\documentclass[structabstract]{aa}
\usepackage[pdftex]{graphics}
\usepackage{graphicx}
\usepackage{txfonts}
\usepackage{lscape}
\usepackage{rotating}
\usepackage[round]{natbib}
\newcommand{\myarcsec}{\hbox{$.\!\!^{\prime\prime}$}}

\newcommand{\be}{\begin{equation}}
\newcommand{\ee}{\end{equation}}

\newcommand{\gabods}{\tt GaBoDS \rm}

\begin{document}
   \title{ARCRAIDER II: Arc search in a sample of non-Abell clusters\thanks{Based on observations made with ESO Telescopes at the La Silla or Paranal Observatories under programme IDs 60.A-9123(G), 65.O-0425, 67.A-0444(A), 067.A-0095(B), 67.A-0427(A), 68.A-0255(A), 69.A-0010(A), 169.A-0595(G), 072.A-0083(A), and 073.A-0050(A). Also based on observations made with the NASA/ESA Hubble Space Telescope, and obtained from the Hubble Legacy Archive, which is a collaboration between the Space Telescope Science Institute (STScI/NASA), the Space Telescope European Coordinating Facility (ST-ECF/ESA) and the Canadian Astronomy Data Centre (CADC/NRC/CSA).}}

   \titlerunning{Arc Search in a Sample of Non-Abell Clusters}

   \author{W. Kausch\inst{1}
      \and
      S. Schindler\inst{1}
      \and
      T. Erben\inst{2}
      \and
      J. Wambsganss\inst{3}
      \and
      A. Schwope\inst{4}
          }

   \offprints{W. Kausch}

   \institute{Institut f\"ur Astrophysik, University of Innsbruck,
             Technikerstr. 25, A-6020 Innsbruck, Austria\\
             \email{wolfgang.kausch@uibk.ac.at}
         \and
             Argelander-Institut f\"ur Astronomie (AIfA), University of Bonn,
         Auf dem H\"ugel 71, D-53121 Bonn, Germany
     \and
         Astrophysikalisches Rechen-Institut und Universit\"at Heidelberg, M\"onchhofstr. 12-14, D-16120 Heidelberg, Germany
     \and
         Astrophysikalisches Institut Potsdam (AIP), An der Sternwarte 16, D-14482 Potsdam, Germany
             }
   \date{Received ; accepted }

% \abstract{}{}{}{}{}
% 5 {} token are mandatory

  \abstract
   {}
   {We present a search for gravitational arcs in a sample of X-ray luminous, medium redshift clusters of galaxies.}
   {The sample of clusters is called ARCRAIDER, is based on the ROSAT Bright Survey (RBS) and fulfills the following criteria: (a) X-ray luminosity
   $\geq0.5\times10^{45}$\,erg/s (0.5-2\,keV band), (b) redshift range $0.1\leq z \leq 0.52$, (c) classified as clusters in the RBS, (d) not a member of the Abell catalogue and, finally, (e) visible from the ESO sites La Silla/Paranal (declination $\delta\leq20^\circ$).}
   {In total we found more than 35 (giant) arc/arclet candidates, including a possible radial arc, one galaxy-galaxy lensing event and a possible quasar triple image in 14 of the 21 clusters of galaxies. Hence 66\% of the sample members are possible lenses. }
   {}


\maketitle
%
%------------------------------------------------------------------------------------------------------------
\section{Introduction}
%------------------------------------------------------------------------------------------------------------
Gravitational lensing techniques have become a blooming branch in
astrophysics in the past with wide ranges of applications. In
particular, the existence of strongly lensed objects offers the
possibility to study the lenses, to investigate high redshift objects
in more detail and can even be used for cosmological researches.
Therefore systematic searches for gravitational arcs may provide an
invaluable basis for those studies. Successful arc searches were
already carried out by several authors, e.g. \cite{bolton08a},
\cite{hennawi08a}, \cite{estrada07a}, \cite{sand05a}, \cite{luppino99a}, \cite{lefevre94a}, \cite{smail91a} or \cite{lynds89a}.\\
A particular application for systematic arc searches is the
statistical approach to arc frequencies, called arc statistics. Arc
statistics investigate the probability of lensing events of
specified properties. These probabilities depend on a large number
of factors, e.g. the number density of sources and lenses, their
properties, and the cosmological model.\\
Among the first to carry out arc statistic simulations in a
cosmological context were \citet[][henceforth
B98]{bartelmann98a,bartelmann03a}. They compared the frequency of
arcs occurring in different cosmological models. Their main result
was that the predicted number of arcs varies by orders of magnitudes
between different cosmologies. In particular, the predicted frequency
of arcs in the currently favoured $\Lambda$CDM model is about one order of
magnitude too low compared to the estimated arc counts derived from
observations. This led to lively discussions on the reasons for that
discrepancy, as the $\Lambda$CDM cosmology is widely supported by different
observations, for example Type I supernovae
\citep{riess98a,perlmutter99a}, or cosmic microwave background
observations \citep[see e.g.][]{hanany00a,pryke02a,spergel03a}.
Therefore arc statistics simulations were refined by several authors:\\
\cite{flores00a} and \cite{meneghetti00a} investigated with
different methods whether contributions of individual cluster
galaxies enlarge the cross section significantly. However, both
found that cluster members do not increase the arc frequency
significantly \citep[$\lesssim15\%$, ][]{flores00a}. Additional effects of source ellipticities and sizes were investigated in detail by \cite{keeton01a}, \cite{oguri02a} and \cite{oguri03a}.\\
The predictions on the lensing efficiency of simulated $0.2\leq
z\leq0.6$ clusters performed by \cite{dalal04a} agree very well with observations based on the Einstein Medium Sensitivity Survey
\citep[henceforth EMSS, ][]{luppino99a}. They also found a strong dependency of about one order of magnitude of the cross section on the viewing angle of the
cluster, which is caused by triaxiality and shallow density cusps of
their simulated clusters.\\
While B98 assumed a constant source distance of
$z_{\rm s}=1$ \citet{wambsganss04a} showed that the
lensing probability is a strong function of the source redshift,
which was confirmed by \cite{li05a}. Varying the source redshift
yields a much higher optical depth, hence the predicted arc
frequency is significantly higher. Using only $z_{\rm s}=1$ sources
\citet{wambsganss04a} confirm the results of B98. \cite{torri04a} investigated the
influence of the dynamical state of galaxy clusters on arc
statistics. They revealed that during merger processes the caustics
change significantly, increasing the number of long and thin arcs by
one order of magnitude. Another factor was introduced by
\cite{puchwein05a}: they investigated the influence of the
intracluster gas and its properties on the lensing efficiency
and find a considerable impact under certain physical conditions.
In particular, cooling and star formation processes may contribute
to the lensing cross section by steepening the mass profile.\\
Numerical simulations nowadays show that the number of arcs for a $\Lambda$CDM
cosmology roughly agrees with observations. However, a direct comparison is hardly possible as all simulations are based on idealised situations. In
particular observers have to deal with observational effects like seeing,
limiting magnitude, instrumental properties,.... which are not taken
into account in the simulations at all. The reason is that those
observational effects are not yet modelled properly, as they concern
a wide range of different effects. For example, observations are
usually based on a set of individual images, which are stacked on
one single final frame. All images are unavoidably taken under
slightly different conditions, hence the final frame contains a
mixture of all individual image properties. The final effect of such
a mixture is hard to judge. In particular, blurring due to different
seeing conditions may affect the length-to-width ratio as well as
the length of an arc, leading to different morphology detections.
The first attempt taking observational effects into account was done
by \cite{horesh05a}. They compared a sample of ten galaxy clusters
based on HST observations \citep{smith05a} with simulations
including some observational effects. Although the observed sample
is very small and based on Abell clusters only, they found an agreement
between arc frequency predictions for $\Lambda$CDM cosmology and the used
observations.\\
We present a sample of galaxy clusters, which is particularly aimed at arc statistics. In Sect.\ref{sec:arcraider} we first describe the cluster sample in detail, its selection criteria, the observations and the data treatment, followed by a report on the methods used
(Sect.\ref{sec:methods}). The results of the arc search are
presented in Sect.\ref{sec:results}, a summary and conclusion is
given in Sect.\ref{sec:summary}. Throughout this paper we use a
standard $\Lambda$CDM cosmology with $H_{\rm
0}=70h_{70}^{-1}$\,km\,s$^{-1}$\,Mpc$^{-1}$, $\Omega_{\rm M}=0.3$,
and $\Omega_\Lambda=0.7$.\\
%------------------------------------------------------------------------------------------------------------
\section{The ARCRAIDER project}\label{sec:arcraider}
%------------------------------------------------------------------------------------------------------------
\subsection{Selection criteria}\label{subsec:selection}
%------------------------------------------------------------------------------------------------------------
ARCRAIDER stands for \bf ARC\rm statistics with X-\bf RA\rm y lum\bf
I\rm nous me\bf D\rm ium r\bf E\rm dhift galaxy cluste\bf R\rm s and
is an ongoing long term project. It is based on the ROSAT Bright
Survey \citep[henceforth RBS]{schwope} and aimed at arc statistic
studies. The RBS is a compilation of the brightest sources in the
ROSAT All Sky Survey \citep{voges} with high galactic latitudes
($|b|>30^\circ$) and a PSPC count rate of $>0.2$\,s$^{-1}$. From this
sample we selected objects fulfilling the criteria
\begin{itemize}
\item classified as a cluster of galaxies
\item a redshift of $0.1\leq z \leq0.52$
\item X-ray luminosity $\log(L_{\rm X})\geq0.5\times10^{45}$\,erg/s (0.5-2\,keV band)
\item visible from the ESO sites La
Silla/Paranal (declination $\delta\leq20^\circ$)
\item not a member of the Abell catalogue \citep{abell}
\end{itemize}
As the resulting 21 clusters (see Table\,\ref{tab:sample} for more details) are located at high galactic latitudes, the $n_{\rm H}$ values are very small ($n_{\rm H}\leq7.7\times10^{20}$cm$^{-2}$). The members of the Abell catalogue were excluded at this first stage of the project as they are also frequent targets of observations during finished and ongoing studies. Hence it is not necessary to reobserve them which reduces our need for observing time.
%---------------------------------------------------------------------------------
%\begin{landscape}
\begin{table*}[Ht]
{\footnotesize
\begin{center}
\caption{Overview of the sample: $L_{\rm X}$ is the X-ray luminosity in the 0.5-2.0\,keV band, computed as $L_{\rm X}=\log\left[4\pi f_x(cz/H_0)^2\right]$, with $H_0=50$kms$^{-1}$Mpc$^{-1}$. All data were taken from
\cite{schwope}.}
\label{tab:sample} \centering
\begin{tabular}{c c c c c c c l }     % 7 columns
\hline\hline
RBS & $\alpha$ (J2000) & $\delta$ (J2000) &  &  & & $n_{\rm H}$ & alternative   \\
number & [h m s] & [d m s] & $z$ & 1\myarcsec0=[kpc] & $L_{\rm X}$ & [$10^{22}\,e^-$/cm$^2$] & name\\
\hline
   RBS-0172 & 01 15 48.8 & -56 55 29 & 0.272 & 4.16 & 44.8 & 3.39 & APMUKS(BJ) B011348.12-571116.8\\
   RBS-0238 & 01 45 11.9 & -60 33 45 & 0.1795 & 3.03 & 44.6 & 3.36 & \\
   RBS-0312 & 02 24 36.4 & -24 33 44 & 0.305 & 4.51 & 44.8 & 1.69 & \\
   RBS-0325 & 02 32 16.4 & -44 20 48 & 0.282 & 4.27 & 44.8 & 2.61 & \\
   RBS-0380 & 03 01 07.5 & -47 06 25 & 0.515 & 6.20 & 45.3 & 2.23 & \\
   RBS-0381 & 03 01 38.5 & +01 55 16 & 0.1690 & 2.88 & 44.5 & 7.7 & ZwCl 0258.9+0142, RHS 18, 4C +01.06 \\
   RBS-0436 & 03 31 06 & -21 00 34 & 0.189 & 3.16 & 44.6 & 2.47 & \\
   RBS-0464 & 03 42 54.2 & -37 07 39 & 0.201 & 3.31 & 44.5  & 1.64& \\
   RBS-0651 & 05 28 15.1 & -29 43 03 & 0.157 & 2.72 & 44.2  & 1.83 & [BCT2000] J052815.92-294300.8 \\
   RBS-0653 & 05 28 52.7& -39 28 18 & 0.286 & 4.31 & 44.9 & 2.1 & NVSS J052853-392815 \\
   RBS-0745 & 09 09 00.4 & +10 59 35 & 0.1600 & 2.76 & 44.5 & 3.91 & MS 0906.3+1111\\
   RBS-0864 & 10 23 39.6 & +04 11 10 & 0.2906 & 4.36 & 45.3  & 2.87 & ZwCl 1021.0+0426, Z3146 \\
   RBS-1015 & 11 40 23.54 & +15 28 10 & 0.24 & 3.79 & 44.9  & 2.65 & \\
   RBS-1029 & 11 45 35.1 & -03 40 02 & 0.1683 & 2.87 & 44.6  & 2.49 & SDSS J114535.10-034001.6\\
   RBS-1267 & 13 26 17.7 & 12 29 58 & 0.2034 & 3.34 & 44.7 & 1.95 & [VCV2001] J132617.6+123000\\
   RBS-1316 & 13 47 32 & -11 45 42 & 0.451 & 5.77 & 45.5 & 4.92 &  RX J1347.5-1145, LCDCS 0829\\
   RBS-1460 & 15 04 07.6 & -02 48 17 & 0.2169 & 3.51 & 45.2  & 5.98 & LCRS B150131.5-023636, QUEST J1504-0248\\
   RBS-1691 & 20 41 50.1 & -37 33 39 & 0.100 & 1.84 & 44.2 & 3.61 & \\
   RBS-1712 & 21 02 04.4 & -24 33 58 & 0.1880 & 3.14 & 44.6  & 5.33 & EXO 2059-247 ID\\
   RBS-1748 & 21 29 39.7 & +00 05 18 & 0.233 & 3.71 & 45.0  & 4.22 & \\
   RBS-1842 & 22 16 56.7 & -17 25 27 & 0.1360 & 2.41 & 44.2  & 2.28 & [BCT2000] J221657.46-172528.3\\

\hline
\end{tabular}
\end{center}
}
\end{table*}
%------------------------------------------------------------------------------------------------------------
\subsection{Observations}\label{subsec:observations}
\begin{figure*}[ht]
   \centering
   \includegraphics[angle=0,width=\hsize]{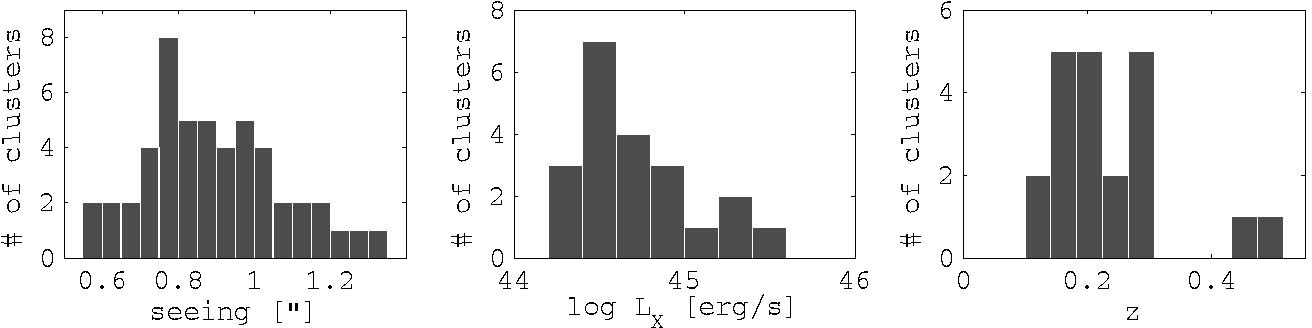}
      \caption{{\it Left}: Seeing histogram. As arc(lets) are often faint and thin structures, it is important to observe under good seeing conditions. The majority of our observations were performed with a seeing better than 1\arcsec, the median value is 0\myarcsec87. {\it Middle:} histogram of the $L_{\rm X}$ distribution of the selected clusters. {\rm Right:} redshift distribution of the sample members.}
         \label{fig:seeinghist}
\end{figure*}

%------------------------------------------------------------------------------------------------------------
The total number of the sample is 21 galaxy clusters, which were usually
observed under good seeing condition (median seeing value
0.87\arcsec, see Figure\,\ref{fig:seeinghist}) at different ESO
telescopes (see Table\,\ref{tab:observations} for more details on the used data, filters and proposal-IDs). As the main instrument we chose SUSI2@NTT, except for RBS325, RBS653 and RBS864. These three clusters were observed with WFI@2.2m, because we had to shift to this telescope during the scheduling process. All clusters were observed at least in the $V$ and the $R$ band to achieve colour information. Additional observations with different filters or instruments were used when available in the ESO archive (see Tables\,\ref{tab:observations} and \ref{fig:limmag}). The given limiting magnitude in Table\,\ref{fig:limmag} is defined as $m_{\rm lim}=ZP-2.5\log(\sqrt{N_{\rm pix}}\cdot 3\cdot\sigma)$, where $ZP$ is the magnitude zeropoint, $N_{\rm pix}$ is the number of image
pixels in a circle with a radius of $2\myarcsec 0$ and $\sigma$ is the sky
background noise.\\
The data reduction was performed with the help of the \gabods pipeline. This software package was especially designed for multi-chip imagers and performs the basic reduction, superflatting and fringe correction, astrometric and photometric calibration and, finally, the
coaddition. For more details on the used algorithms we refer the
interested reader to \cite{pipeline1} and \cite{pipeline2}.\\
The astrometric reference frame was tied to the USNO-A2 catalogue
\citep{mon98,monet98a}, the photometric calibration was done with the STETSON standards \citep{stetson00a}. All magnitudes are given in the Vega system.\\
For nights where no standard star observations were observed we took the standard zero points given on the SUSI2 homepage\footnote{\tt\scriptsize
http://www.ls.eso.org/lasilla/sciops/ntt/susi/docs/SUSIphot.html} after investigating the photometric conditions of these nights with the help of WFI observations.\\
%--------------------------------------------------------------------------
\begin{table*}[Ht]
\begin{footnotesize}
\caption{Overview of the observations; the instrument mode
describes the chosen binning mode for SUSI2, the resolution
mode for FORS, and the imaging mode for VIMOS. The used WFI
filters were Ic/Iwp (ESO845), R$_{\rm c}$/162 (ESO844), V/89 (ESO843), and, for RBS-0864, B/123 (ESO878), whereas B/99 (ESO842) was used for RBS-0325 and RBS-0653, respectively. The used SUSI2 Bessel filters were V\#812, R\#813, I\#814, the FORS filters U\_BESS+33, B\_BESS+34, V\_BESS+35, R\_BESS+36, and I\_BESS+37. ($^1$ see \citet{kausch07a}, $^2$ see \citet{bradac05a,halkola08a}, $^3$ filter ACS/WFPC2-$F606W$, $^4$ Hubble proposal ID.)}
\label{tab:observations} \centering
\begin{tabular}{c | c c | c c c c c c | l }     % 7 columns
\hline\hline
RBS & instrument & instrument &$U$ & $B$ & $V$ & $R$ & $I$ & $Ks$ & Proposal\\
number & & mode & $t_{exp} [s]$ & $t_{exp} [s]$ & $t_{exp} [s]$ & $t_{exp} [s]$ & $t_{exp} [s]$ & $t_{exp} [s]$ & IDs\\
\hline
   RBS-0172 & SUSI2@ESONTT & $2\!\times\!2$ & - & - & 3040 & 6080 & - & - & 67.A-0444(A)\\
   RBS-0238 & SUSI2@ESONTT & $2\!\times\!2$& - & - & 3040 & 6080  & - & - & 67.A-0444(A)\\
   RBS-0312 & SUSI2@ESONTT & $2\!\times\!2$& - & - & 3040 & 6080  & - & - & 67.A-0444(A)\\
   RBS-0325 & WFI@ESO2.2m & - & - & 5400 & 8000 & 16100 & - & - & 68.A-0255(A)\\
   RBS-0380 & SUSI2@ESONTT & $2\!\times\!2$& - & - & 2250 & 2250 & - & - & 67.A-0444(A)\\
   RBS-0381 & SUSI2@ESONTT & $1\!\times\!1$ & - & - & 2600 & 5200 & - & - & 072.A-0083(A)\\
   RBS-0436 & SUSI2@ESONTT & $1\!\times\!1$ & - & - & 2600 & 5200 & - & - & 072.A-0083(A)\\
            & ACS@HST  & - & - & - & 1200$^3$ & - & - & - & 10881$^4$\\
   RBS-0464 & SUSI2@ESONTT & $1\!\times\!1$ & - & - & 3600 & 5900 & - & - & 072.A-0083(A)\\
   RBS-0651 & SUSI2@ESONTT & $1\!\times\!1$ & - & - & 1740 & 5200 & - & - & 072.A-0083(A)\\
            & ACS@HST & - & - & - & 1200$^3$ & - & - & - & 10881$^4$\\
   RBS-0653 & WFI/ESO2.2m & - & - & 1200 & 8000 & 8000 & 1200 & - & 68.A-0255(A), 60.A-9123(G)\\
   RBS-0653 & VIMOS/IMG & - & - & - & - & 2400 & - & - & 169.A-0595(G)\\
   RBS-0745 & SUSI2@ESONTT & $2\!\times\!2$& - & - & 3040 & 6080  & - & - & 69.A-0010(A)\\
   RBS-0864$^1$ & WFI@ESO2.2m & -& - & 1500 & 8000 & 25000 & - & - & 68.A-0255(A), 073.A-0050\\
   RBS-1015 & SUSI2@ESONTT & $1\!\times\!1$ & - & - & 2600 & 5900 & - & - & 072.A-0083(A)\\
   RBS-1029 & SUSI2@ESONTT & $2\!\times\!2$& - & - & 3040 & 6080  & - & - & 69.A-0010(A)\\
   RBS-1267 & SUSI2@ESONTT & $1\!\times\!1$ & - & - & 2200 & 6080 & - & - & 072.A-0083(A)\\
   RBS-1316$^2$ & FORS/ISAAC@VLT& High Res./- &11310 & 4800 & 4500 & 6000 & 6750 & $\sim7200$  & 67.A-0427(A), 067.A-0095(B) \\
   RBS-1460 & SUSI2@ESONTT & $2\!\times\!2$& - & - &  3040 & 6080 & - & -  & 69.A-0010(A)\\
   RBS-1691 & SUSI2@ESONTT & $2\!\times\!2$& - & - & 3040 & 6080  & - & - & 69.A-0010(A)\\
   RBS-1712 & SUSI2@ESONTT & $2\!\times\!2$& - & - & 15380 & 6080 & 9300 & - & 65.O-0425, 67.A-0444(A)\\
   RBS-1748 & SUSI2@ESONTT & $2\!\times\!2$& - & - & 3040 & 6080  & - & - & 67.A-0444(A)\\
            & WFPC2@HST & - & - & - & 1200$^3$ & -  & - & - & 8301$^4$\\
   RBS-1842 & SUSI2@ESONTT & $2\!\times\!2$& - & - & 3800 & 7600 & - & - & 69.A-0010(A)\\
%   RBS1971 & SUSI2@ESONTT & $2\!\times\!2$& - & - & 11300 & 6000 & - & - & \\
\hline
\end{tabular}
\end{footnotesize}
\end{table*}
%--------------------------------------------------------------------------
%--------------------------------------------------------------------------
Where available we also used HST data taken from the archive. This concerns the clusters RBS-0436$^a$, RBS-0651$^{\rm a}$,RBS-0864$^{\rm b}$ and RBS-1748$^{\rm b}$, which were observed during snapshot programmes (filter: $F606W$; $^{\rm a}$PID: 10881, P.I. Smith, ACS, $t_{\rm exp}=1200$\,s; $^{\rm b}$PID: 8301, P.I.: Edge, WFPC2, $t_{\rm exp}=1000$\,s). We used the calibrated images from the Hubble Legacy Archive\footnote{\tt\scriptsize http://hla.stsci.edu/} mainly to identify possible arcs, as those images are well-suited due to their missing atmospheric blurring, even in spite of the very short exposure times.
%--------------------------------------------------------------------------
\begin{table}[h]
{\footnotesize \centering \caption{Table with measured limiting magnitude values. \label{fig:limmag}}
\begin{tabular}{c | c c c c }
\hline\hline
RBS     & m$_{\rm lim}$ & m$_{\rm lim}$ & m$_{\rm lim}$ & m$_{\rm lim}$ \\
cluster & in $B$ [mag]  &  in $V$ [mag]  &  in $R$ [mag]  &  in $I$ [mag] \\
\hline
RBS-0172  & -- & 25.67 & 25.73 & --  \\
RBS-0238  & -- & 25.63 & 25.57 & --  \\
RBS-0312  & -- & 25.66 & 25.55 & --  \\
RBS-0325  & 25.81 & 25.46 & 25.49 & --  \\
RBS-0380  & -- & 25.22 & 25.44 & --  \\
RBS-0381  & -- & 25.95 & 25.74 & --  \\
RBS-0436  & -- & 25.71 & 25.29 & --  \\
RBS-0464  & -- & 25.72 & 25.69 & --  \\
RBS-0651  & -- & 25.27 & 25.80 & --  \\
RBS-0653  & 24.89 & 25.07 & 25.76 & 22.35\\
RBS-0745  & -- & 25.13 & 25.18 & -- \\
RBS-1015  & -- & 26.19 & 25.80 & --  \\
RBS-1029  & -- & 25.30 & 25.21 & --  \\
RBS-1267  & -- & 25.86 & 24.40 & --  \\
RBS-1460  & -- & 25.12 & 25.34 & --  \\
RBS-1691  & -- & 25.41 & 25.48 & --  \\
RBS-1712  & -- & 25.19 & 25.91 & 24.70\\
RBS-1748  & -- & 25.51 & 25.47 & --  \\
RBS-1842  & -- & 25.71 & 25.55 & --  \\
\hline
\end{tabular} }
\end{table}
%--------------------------------------------------------------------------
%------------------------------------------------------------------------------------------------------------
\section{Methods}\label{sec:methods}
%------------------------------------------------------------------------------------------------------------
\subsection{Determination and photometry of the arc candidates}\label{subsec:arcdetermination}
%------------------------------------------------------------------------------------------------------------
One of the main issues of the ARCRAIDER project at this stage is the search for gravitational arcs. As arcs are difficult to detect, we have to focus on the most important criteria. In ground-based observations usually only arcs
tangentially aligned with respect to the mass centre are visible.
Radial arcs are often too thin and too faint structures in the
vicinity of bright central galaxies of clusters. In addition, arcs
and their counter images have the same spectra and redshifts of the order of
$\gtrsim2\times z_{\rm lens}$. However, as we do not have spectra, we
restrict our search criteria to the morphology, the position
and alignment of the possible candidates with respect to a bright central cluster galaxy (BCG), assuming the latter to be the centre.\\
To determine the length-to-width ratio $l/w$ to be used as selection criterion we follow an ansatz by \cite{lenzen} and
the \tt SExtractor \rm Manual \citep{sextractor} and define the $l/w$ ratio by
calculating the eigenvalues $\lambda_1$ and $\lambda_2$ of the second order
moment of the light distribution $L_{kl}$ (note that $L_{12}=L_{21}$):
\begin{eqnarray}\label{eq.l_to_w}
    \lambda_1^2=\frac{L_{11}+L_{22}}{2}+\sqrt{\left(\frac{L_{11}-L_{22}}{2}\right)^2+L_{12}}\\
    \lambda_2^2=\frac{L_{11}+L_{22}}{2}-\sqrt{\left(\frac{L_{11}-L_{22}}{2}\right)^2+L_{12}},
\end{eqnarray}
the \it ratio \rm $l/w$ is equal to $\lambda_1/\lambda_2$ \citep{jaehne}. Hence we obtain the length-to-width ratio by determining $\lambda_1$ and $\lambda_2$.\\
Though a lot of useful information (e.g. length-to-width ratio,
location/orientation with respect to the BCG...) is contained in the catalogues, it is insufficient to restrict the search to object catalogues only. In several cases arcs merge apparently with foreground objects and can
therefore be missed or are simply too faint to be detected. Hence an
arc search is best performed by visual inspection of deep images and
not restricted to catalogues only. We use the following
selection criteria for arc candidates: (a) they are tangentially aligned,
and, (b) in a distance of $<1\arcmin$, both with respect to the
central cluster galaxy, and (c) show a length-to-width ratio of $l/w\geq1.5$
(measured with SExtractor\rm, see Sect.\,\ref{subsec:arcmorphology} for more details).\\
We additionally assigned the arc candidates to two classes (\tt
A\rm, and \tt B\rm). Class \tt A \rm
denotes a high probability of being a lensed object, whereas \tt B \rm
type objects are of low probability, but not excludable lensing
features. This separation is primarily meant to be as priority list for subsequent observations.\\
All magnitudes were determined with \tt SExtractor \rm in double image mode using MAG\_AUTO  with the following parameters for all clusters: DETECT\_MINAREA=3, effective GAIN$=t_{exp}*{\rm GAIN_{\rm INSTR}}$. The GAIN$_{\rm INSTR}$ is 2 for WFI and 2.25 for SUSI2. We also took
galactic extinction $E(B-V)$ into account based on values by \cite{schlegel}, in spite of the low values due to the high galactic latitude bias introduced by the RBS catalogue. The separation of stars and galaxies was done using CLASS\_STAR $<0.95$.\\
Additional information about the arc candidates can be derived from their colour information. We have compared the $(V-R)$ colours of the arcs with the average colour of the five brightest cluster members (see Table\,\ref{tab:arclist1}). As lensed objects are highly redshifted galaxies of various types, their colour usually differs from the main lensing cluster members. Except for three of them (RBS-0238: \tt B2\rm; RBS-0651: \tt B3\rm; RBS-1460: \tt B1\rm) all candidates show different colours than the five brightest cluster members, which indicates their non-cluster membership.\\
%------------------------------------------------------------------------------------------------------------
\subsection{Morphology of the arc candidates / mass estimates}\label{subsec:arcmorphology}
\begin{figure}[ht]
   \centering
   \includegraphics[angle=0,width=\hsize]{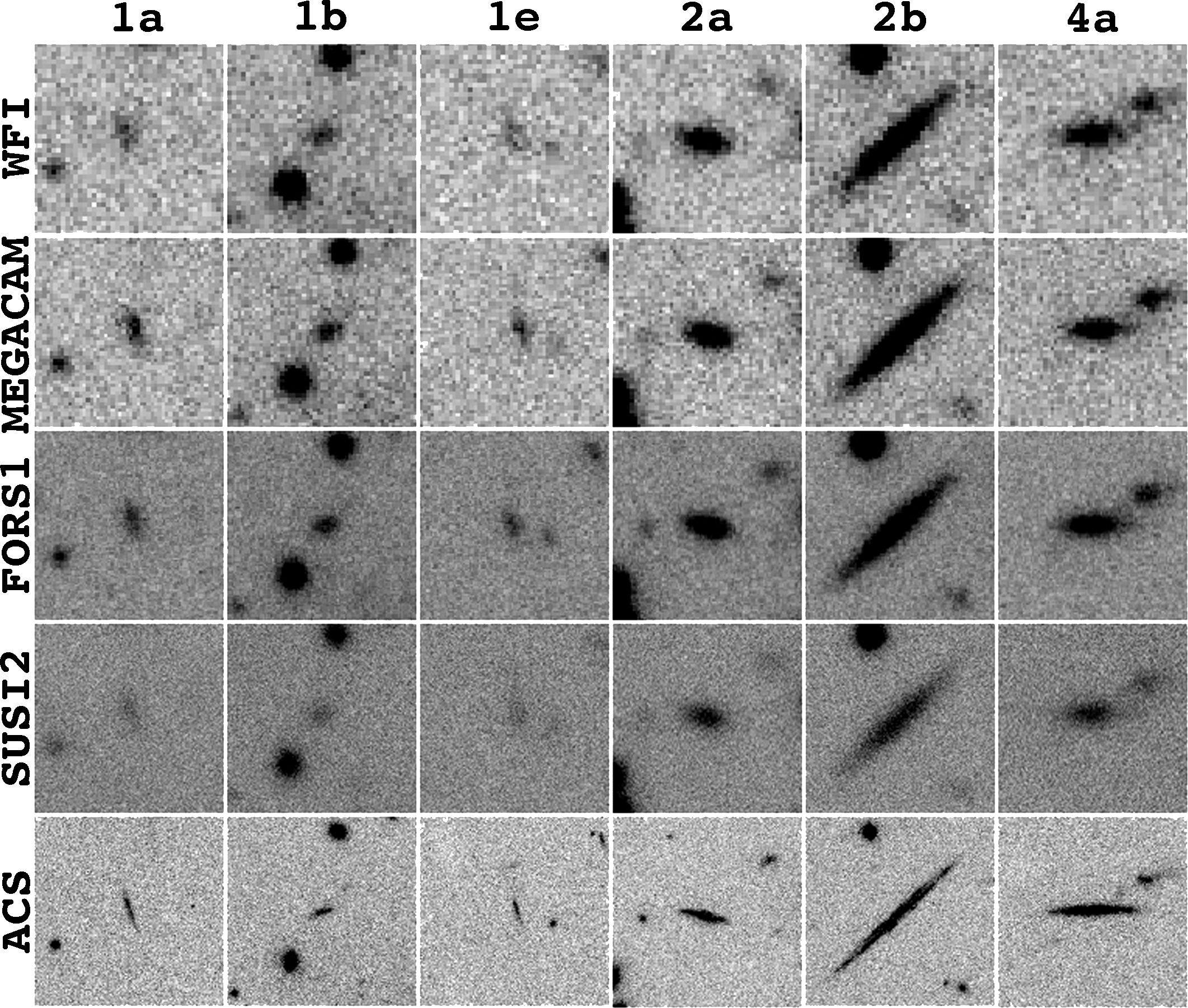}
      \caption{Comparison of various gravitational arcs seen with five different imagers: WFI@ESO2.2m, MEGACAM@CFHT, FORS1@VLT, SUSI2@NTT and ACS onboard the HST. The naming of the arcs is taken from \cite{halkola08a}.
      \label{fig:5cam_comparison}}
\end{figure}

%------------------------------------------------------------------------------------------------------------
At first glance the chosen limit of the length-to-width ratio $l/w\geq1.5$ for arc determination seems to be very low. However, atmospheric blurring dramatically decreases the $l/w$. Figure\,\ref{fig:5cam_comparison} shows $R$-band images of five gravitational arclets and three arcs discovered in RXJ1347-1145 \cite[see][]{bradac05a,halkola08a} as seen by four different ground based imagers (WFI@ESO2.2m, MEGACAM@CFHT, FORS1@ESOVLT, SUSI2@ESONTT) in comparison with $F814W$ frame observations (no $R$ band was available) taken with the ACS onboard the HST (see Table\,\ref{tab:5cam_comparison}). All observations are very deep (mag$_{\rm lim}\leq25.4$) and all ground based images were taken under excellent seeing conditions (seeing $<1\arcsec$). To estimate the influence of $l/w$ measurements caused by atmospheric blurring we define the blurring factor BF as the ratio $(l/w)_{\rm gba} / (l/w)_{\rm ACS}$, (where gba stands for "ground based average").\\
Except for object 1e, which is by far the smallest and faintest, the length-to-width ratio measurements on the ground based images roughly agrees (see Table\,\ref{tab:l_over_w}). However, it is clearly visible that the $l/w$ is dramatically higher in the HST frame, which is shown by a  blurring factor BF of down to $\sim0.4$. This means that the influence of atmospheric blurring is highly dominant over other factors like different spatial sampling (pixel scale), or exposure times. Hence the fraction of missed arc candidates is minimized with an assumed limit of the ground based length-to-width ratio $l/w\geq1.5$ for arc determination. However, the contamination by cluster or foreground elongated objects is increased. Thus a spectroscopic confirmation of the arc candidates is mandatory in the future.\\
We can also use the position of the arcs to roughly estimate the
mass of the lensing galaxy assuming it is a part of an Einstein
ring. As we do not know the distance to the background galaxy we
only get a rough estimate of the mass within the Einstein ring by
assuming $z_{\rm background}=1$, the upper/lower limits are
estimated by assuming $z_{\rm background}=[2\times z_{\rm cluster},
2]$. Due to the large number of assumptions we concentrate on \tt A \rm class candidates only.
%--------------------------------------------------------------------------
%--------------------------------------------------------------------------
%--------------------------------------------------------------------------
\begin{table}[ht]
\begin{center}
{\footnotesize \centering \caption{Technical data; used filters: $R_{\rm c}162$ (WFI), SDSS-$r$ (MEGACAM),
R\_BESS+36 (FORS1), WB655\#825 (SUSI2), and F814W (ACS)\label{tab:5cam_comparison}}
\begin{tabular}{c | c c c c}
\hline\hline
Imager     & pixel & measured          &   exposure  & m$_{\rm lim}$ \\
           & scale & seeing [\arcsec]  &    time [s] &  [mag]    \\
\hline
 WFI     & 0.238 & 0.85 &   16300 & 25.4  \\
 MEGACAM & 0.186 & 0.75 &   7200  & 25.9 \\
 FORS1   & 0.1   & 0.69 &   6000  & 26.5 \\
 SUSI2   & 0.0805& 0.91 &   11250 & 25.9 \\
 ACS     & 0.05  & n/a  &   5280  & 26.45 \\
\hline
\end{tabular} }
\end{center}
\end{table}
%--------------------------------------------------------------------------
%--------------------------------------------------------------------------
\begin{table}[ht]
\begin{center}
{\footnotesize \centering \caption{Measured $l/w$ of the arcs shown in Figure\,\ref{fig:5cam_comparison}. \label{tab:l_over_w}}
\begin{tabular}{c | c c c c c c}
\hline\hline
Imager     & 1a & 1b & 1e  & 2a & 2b & 4a \\
\hline
 WFI     & 1.6 & 1.5 & 1.1 & 1.7 & 4.5 & 2.5\\
 MEGACAM & 2.0 & 1.5 & 1.9 & 1.8 & 5.0 & 2.5\\
 FORS1   & 1.8 & 1.4 & 1.7 & 1.5 & 4.9 & 2.6\\
 SUSI2   & 1.8 & 1.3 & 1.4 & 1.7 & 4.7 & 2.4\\
 ACS     & 2.6 & 3.1 & 3.8 & 4.3 & 8.1 & 7.4\\
\hline
BF & $0.69$ & $0.46$ & $0.40$ & $0.39$ & $0.59$ & $0.34$ \\
\hline
\end{tabular} }
\end{center}
\end{table}
%--------------------------------------------------------------------------
%------------------------------------------------------------------------------------------------------------
\section{Results and discussion}\label{sec:results}
%------------------------------------------------------------------------------------------------------------
%------------------------------------------------------------------------------------------------------------
\subsection{Individual clusters}\label{subsec:individual}
%------------------------------------------------------------------------------------------------------------
The ARCRAIDER sample includes several clusters with distinct arc like features (see Table\,\ref{tab:arclist1} and the Appendix$^3$ for a complete list and images). In particular RBS-0325, RBS-0651, RBS-0653, RBS-0864 \citep{kausch07a} and RBS-1316 \cite[RXJ1347-1145, see e.g.][]{bradac05a,halkola08a} show distinct strong lensing candidates. Apart from that several small arclet candidates, the \tt B\rm-typed objects, can be found in various clusters of the sample, however their lensing probability is very low. In addition we found a galaxy-galaxy lensing candidate in RBS-0312, a candidate for a radial arc  in RBS-0325 (see Figures\,A.1c and d, respectively\footnote{see download link in the Appendix}), and a possible multi-imaged quasar in RBS-1712 (Figure\,A.3c$^3$).

%------------------------------------------------------------------------------------------------------------
\subsection{Correlation of the X-ray luminosity and the number of arc candidates}\label{subsec:correlation}
%------------------------------------------------------------------------------------------------------------
A correlation between the X-ray luminosity and the number of arcs is
expected, because the X-ray luminosity correlates with the mass of the
cluster \citep{schindler3,reiprich} and the
probability to detect arcs increases with the cluster mass. Dividing the sample into classes with the X-ray luminosity intervals \tt I\rm: [$44.2\leq\log(L_{\rm X})\leq44.5$], \tt II\rm: [$44.5< \log(L_{\rm X})<44.9$], and \tt III\rm: [$44.9\leq\log(L_{\rm X})$] to obtain three classes of approximately similar size, we find a strong correlation between the number of \tt A\rm-type arc candidates and the X-ray luminosity in the preliminary sample with 0.33 arc candidates per cluster in the faintest class and 3.14 arc
candidates per cluster in the most luminous bin (see Table\,\ref{tab:sample_L_X}).\\
Seven of the 14 ARCRAIDER lensing cluster candidates show \tt B\rm -type objects only, and six at least one class \tt A \rm type. Additionally $\sim10\%$ are really impressive lensing clusters (RBS-0653 and RBS-1316), containing a large number of candidates and giant arcs.
%-------------------------------------------------------------------------------
\begin{table}[ht]
\begin{center}
{\footnotesize \centering \caption{Sample divided into three classes \tt I\rm, \tt II\rm, and \tt III\rm, with respect to the X-ray luminosity $L_{\rm X}$. \label{tab:sample_L_X}}
\begin{tabular}{l l | c c c}
\hline
 $L_{\rm X}$ & & \#  & \# of \tt A\rm-type arc \\
 class & & clusters & cand. per cluster& \\
\hline
 \tt I \rm   & [$\log(L_{\rm X})\leq44.5$]    & 6 & 0.33 \\
 \tt II \rm  & [$44.5< \log(L_{\rm X})<44.9$] & 8 & 0.5 \\
 \tt III \rm & [$44.9\leq\log(L_{\rm X})$]    & 7 & 3.14 \\
\hline
\end{tabular} }
\end{center}
\end{table}
%--------------------------------------------------------------------------------
%-------------------------------------------------------------------------------
\subsection{Comparison with the EMSS sample}\label{subsec:RBS_EMSS_comparison}
\begin{figure}[ht]
   \centering
   \includegraphics[angle=0,width=\hsize]{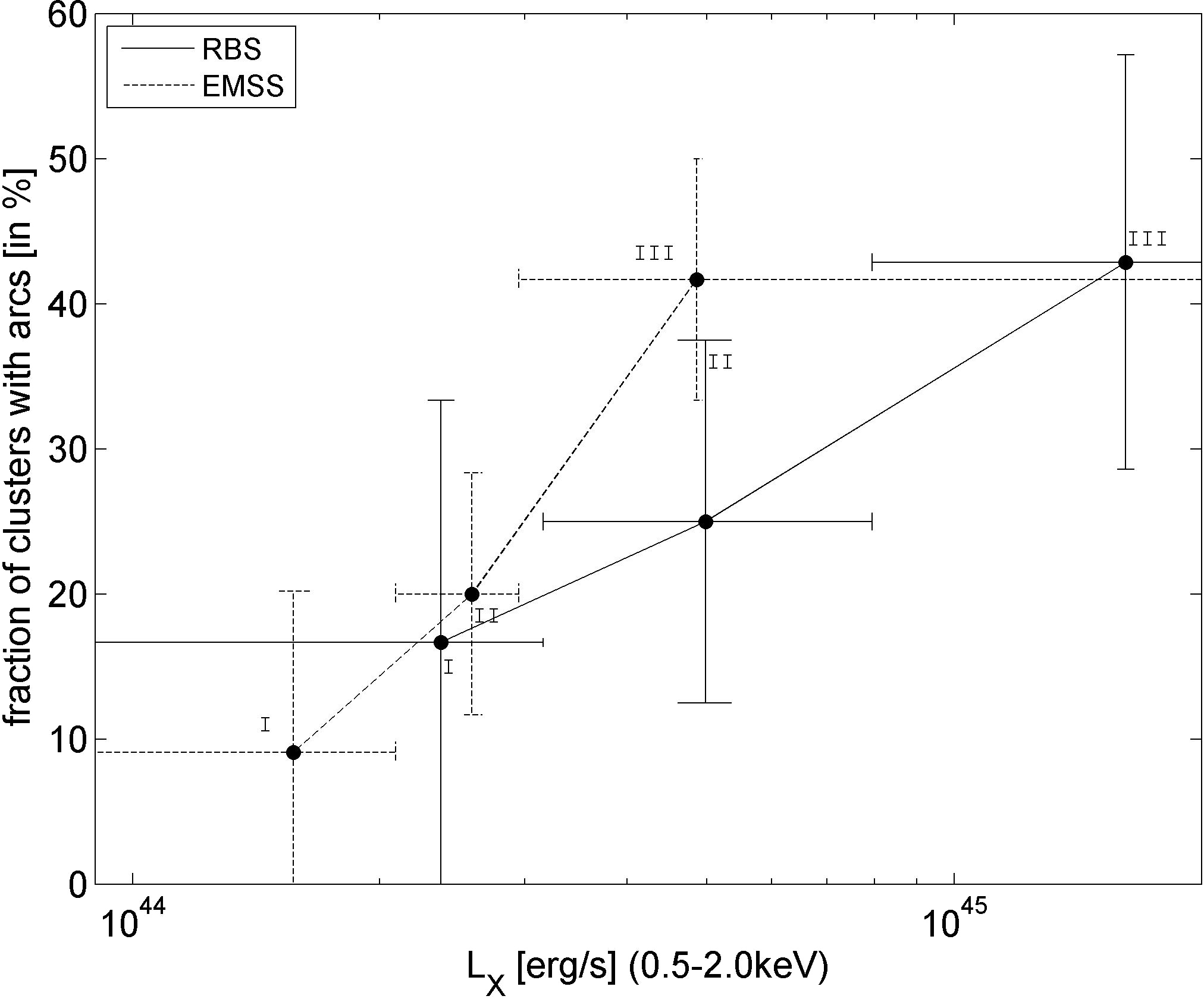}
      \caption{Comparison of the lensing cluster fractions in the EMSS sample (dashed line) \citep{luppino99a} and the RBS sample (solid line). For both samples we used only the most secure arc candidates (see Sect.\,\ref{subsec:RBS_EMSS_comparison}). The dots mark the mean values of the X-ray luminosity within the corresponding $L_{\rm X}$ classes \tt I\rm, \tt II\rm, and \tt III\rm, respectively (see Table\,\ref{tab:sample_means}), error bars in $L_{\rm X}$ mark the limits given in Table\,\ref{tab:sample_L_X}. The error bars in the $y$-direction are assumed to be $\pm1$ cluster with arcs. See Sect.\,\ref{subsec:RBS_EMSS_comparison}
      for more detail.
      \label{fig:RBS_EMSS_comparison}}
\end{figure}

%-------------------------------------------------------------------------------
Several other arc searches have been carried out by various groups \cite[e.g.][]{bolton08a,hennawi08a,estrada07a,sand05a}. However, a direct comparison with those studies is hampered by considerable differences between the studies e.g. in the selection criteria of the cluster samples, photometric depth, or chosen instruments. \\
The only comparable search for gravitational arcs in clusters of galaxies was performed by \cite{luppino99a}, who searched for strong lensing features in 38 X-ray selected clusters taken from the EMSS \citep{gioia90a,stocke91a}. In total they discovered 16 clusters with arcs and arc candidates, including eight systems with giant arcs. 60\% of their clusters exceeding $L_{\rm X}>10^{45}$\,erg\,s$^{-1}$ ($0.3-3.5$\,keV band) inhabitate giant arcs and none of the 15 clusters with $L_{\rm X}<4\cdot10^{44}$\,erg\,s$^{-1}$ (same band) shows any strong lensing feature candidate.\\
Due to the similarities in the samples we can roughly compare the fraction of clusters inhabitating gravitational arcs between both samples. Because of the uncertainties in the arc determination we only take clusters into account with high probability strong lensing features. Hence, we use only type \tt A \rm arcs for our RBS sample (see Sect.\,\ref{subsec:arcmorphology}), for the EMSS sample we use giant arcs and arcs without a question mark in the last column of Table 1 in \cite{luppino99a} and additionally only select clusters with a comparable redshift ($z\leq0.515$).\\
For this comparison we also divided the EMSS sample members into three classes of similar number counts with respect to the X-ray luminosity (see Table\,\ref{tab:EMSS_LX_class}), which was transferred to the $0.5-2.0$\,keV band with the online PIMMS-Tool\footnote{\tt\scriptsize http://heasarc.nasa.gov/Tools/w3pimms.html}.\\
%--------------------------------------------------------------------------------
\begin{table}[ht]
\begin{center}
{\footnotesize \centering \caption{X-ray luminosity classes of the EMSS Sample. \label{tab:EMSS_LX_class}}
\begin{tabular}{l l | c  }
\hline
 $L_{\rm X}$ [$0.5-2$\,keV] & & \# \\ % & \# of  & \# of arcs\\
 EMSS sample class & & clusters  \\ % & arcs & per cluster\\
\hline
 \tt I \rm   & [$\log(L_{\rm X})\leq44.32$]    & 11 \\
 \tt II \rm  & [$44.32< \log(L_{\rm X})<44.47$] & 10 \\
 \tt III \rm & [$44.47\leq\log(L_{\rm X})$]    & 12 \\
\hline
\end{tabular} }
\end{center}
\end{table}
%--------------------------------------------------------------------------------
Figure\,\ref{fig:RBS_EMSS_comparison} shows a comparison of the lensing cluster fraction in the EMSS and the RBS sample, respectively. The dots mark the mean value of the X-ray luminosity of the clusters within the class (see Tables\,\ref{tab:sample_L_X} and \ref{tab:EMSS_LX_class}) of the corresponding sample (see Table\,\ref{tab:sample_means}). The error bars in $x$-direction denote the limits of the corresponing $L_{\rm X}$ class, in the $y$-direction the errors are assumed to be $\pm1$ cluster with missed or misinterpreted arcs, respectively.\\
Surprisingly, the agreement between the samples is not as good as one one would expect for two similar samples. Both are strictly selected by $L_{\rm X}$, except that the luminosity cut is much higher in the RBS sample. However, the discrepancy between the two samples could be caused by the momentary skipping of the Abell clusters ($\sim43\%$ of the clusters in class \tt I\rm, $\sim63\%$ in class \tt II\rm, and $\sim82\%$ in class \tt III\rm). Although the Abell clusters are selected by visible light luminosity only, the omitting excludes several famous prominent lensing clusters \citep[Abell 2204, Abell 2667, Abell 1835, Abell 1689, see e.g.][]{sand05a,broadhurst05a}.\rm
%--------------------------------------------------------------------------------
\begin{table}[ht]
\begin{center}
{\footnotesize \centering \caption{Mean values of the EMSS- and RBS-cluster X-ray luminosity corresponding to the $L_{\rm X}$ ($0.5-2.0$\,keV) classes \tt I\rm, \tt II\rm, and \tt III \rm defined in Tables\,\ref{tab:EMSS_LX_class} and \ref{tab:sample_L_X}, respectively.\label{tab:sample_means}}
\begin{tabular}{c | c c }
\hline
$L_{\rm X}$ & EMSS sample & RBS sample\\
class & $\times10^{45}$\,erg/s & $\times10^{45}$\,erg/s\\
\hline
 \tt I \rm   & $<L_{\rm X}>=0.16$  & $<L_{\rm X}>=0.24$ \\
 \tt II \rm  & $<L_{\rm X>}=0.26$  & $<L_{\rm X}>=0.50$ \\
 \tt III \rm & $<L_{\rm X}>=0.49$  & $<L_{\rm X}>=1.62$ \\
\hline
\end{tabular} }
\end{center}
\end{table}
%--------------------------------------------------------------------------------
%--------------------------------------------------------------------------
%--------------------------------------------------------------------------
%------------------------------------------------------------------------------------------------------------
\section{Summary}\label{sec:summary}
%------------------------------------------------------------------------------------------------------------
We present a systematic search for gravitational arcs in a unique sample of X-ray-luminous, medium redshifted galaxy clusters. The search is based on deep ground based images taken with ESO telescopes under good seeing conditions (usually $<1\arcsec$). Including RBS-0864 \citep[Z3146,][]{kausch07a} and RBS-1316 \citep[RXJ1347-1145,][]{bradac05a, halkola08a}, respectively, the sample consists of 21 members. In total we found candidates for more than 35 (giant) arcs or arclets, one radial arc candidate, one galaxy-galaxy lensing event and three possible quasar lensing features in 14 sample members. Hence $~66\%$ of the clusters are possible strong lenses, with a strong bias towards X-ray luminous clusters (see Sect.\,\ref{subsec:correlation}). \\
The next step in the ARCRAIDER project is the analysis of the currently excluded Abell clusters belonging to this sample and spectroscopic follow-up observations of the arc candidates to confirm their lensing nature. Including these missing clusters and observations the ARCRAIDER sample is by far the largest for future arc statistic studies.
%------------------------------------------------------------------------------------------------------------
\begin{landscape}
\begin{center}
\begin{table}
%\vspace{2.3cm}
%\hspace{3cm}
{\caption{List of the photometric and morphologic
properties of measurable arcs and arc candidates, except for RBS-0864 \citep[see][]{kausch07a} and RBS1316 \citep[RXJ1347-1145, see][]{bradac05a}. The arcs are usually named by their classification (see Sect.\,\ref{subsec:arcdetermination}) followed by a running number. The distances are measured from the cluster centres (hereafter 'cc'), the given masses are estimated as masses within an Einstein radius of the \tt A \rm class candidates with a source redshift $z_s=1$, the errors are calculated with $z_s=2*z_{\rm cluster},2$. As some arcs are very faint structures it was not possible to obtain reliable photometric information or morphologic estimates (marked by '--'); The determination of the length-to-width ratio $l/w$ is described in Sect.\,\ref{subsec:arcmorphology}. Remarks: $^1$ merging with another object;
$^2$ galaxy-galaxy lensing candidate; $^3$ mass of the single galaxy with
GGL1; $^4$ GA=giant arc; $^5$ measured on the HST frame; $^6$ mass of the BCG; $^7$ averaged colour of the five brightest cluster galaxies. \label{tab:arclist1}}
\begin{center}
\vspace{0.7cm}
\begin{tabular}{c | c | c c c c | c c | c c c c }     % 7 columns
\hline\hline
                      % To combine 4 columns into a single one
cluster     & arc   &  $I$  &  $R$  &  $V$  &  $B$  & $<(V-R)>$ & $(V-R)$ & $l/w$ & distance to & distance to & mass\\
            &       & [mag] & [mag] & [mag] & [mag] & cluster$^7$ & arcs &  & cc [\arcsec] & cc [$h^{-1}_{70}$\,kpc] & [$\times 10^{14}{\rm M}_\odot$]\\
\hline
  RBS-0172 & B1    &  --  & $23.28\pm0.02$ & $24.16\pm0.07$ & -- & 0.72 & $0.88\pm0.09$ & 1.5 & $\sim29$ & $\sim120$ & -- \\
\hline
  RBS-0238 & B1    &  --  & $22.42\pm0.04$ & $23.31\pm0.05$ & -- & 0.72 & $0.89\pm0.10$ & 2.5 & $\sim20$ & $\sim60$  & --  \\
            & B2$^1$&  --  & $23.61\pm0.07$ & $24.32\pm0.08$ & -- & 0.72 & $0.70\pm0.15$ & 2.0 & $\sim50$ & $\sim150$ & -- \\
            & B3$^1$&  --  &      --       &      --     & -- & -- & -- &  --  & $\sim20$ & $\sim60$  & --\\
\hline
   RBS-0312 & GGL1 $^2$ & -- & -- & -- & -- & -- & -- & -- & $\sim2.4$ & $\sim11$ & $0.0104^{+0.004}_{-0.002}$  $)^3$ \\
\hline
   RBS-0325 & A1  & -- & $21.02\pm0.01$ & $21.48\pm0.02$ & $21.79\pm0.03$ & 0.84 & $0.47\pm0.02$ & 1.7 & $\sim50$ &$\sim210$ & $4.04^{+1.73}_{-0.64}$\\
          & B2  & -- & $22.58\pm0.01$ & $23.38\pm0.07$ & $24.09\pm0.23$ & 0.84 & $1.80\pm0.08$ & 1.7 & $\sim47$ &$\sim200$ & --\\
          & A3  & -- & $22.89\pm0.03$ & $24.29\pm0.12$ & $24.60\pm0.30$ & 0.84 & $1.40\pm0.15$ & 2.6 & $\sim30$ &$\sim130$ & $1.48^{+0.63}_{-0.23}$\\
          & B4  & -- & -- & -- & -- & -- & -- & -- & $\sim38$ &$\sim162$ & $2.38^{+1.01}_{-0.37}$\\
\hline
  RBS-0436 & B1    &  --  & $23.04\pm0.05$ & $23.47\pm0.01$ & -- & 0.56 & $0.42\pm0.06$ & 2.7 & $\sim41$ &$\sim130$ & -- \\
\hline
  RBS-0651 & A1    &  --  & $21.11\pm0.01$ & $22.31\pm0.05$ & -- & 0.59 & $1.20\pm0.05$ & 2.5 & $\sim25$ & $\sim67$ & $0.52^{+0.35}_{-0.04}$ \\
          & A2 & -- & -- & -- & -- & -- & -- & -- & $\sim12$ & $\sim30$ & $0.123^{+0.08}_{-0.01}$)$^6$\\
          & B3 & -- & $21.38\pm0.01$ & $22.79\pm0.09$ & -- & 0.59 & $0.58\pm0.20$ & 2.8 & $\sim37$ & $\sim100$ & --\\
\hline
   RBS-0653 & GA1$^4$ & -- & -- & -- & -- & -- & -- & -- &  $\sim20$ & $\sim87$  & $6.67^{+2.80}_{-1.07}$ \\
          & GA2 & -- & -- & -- & -- & -- & -- & -- &  $\sim11$ & $\sim47$ & $2.02^{+0.85}_{-0.32}$\\
          & GA3 & -- & -- & -- & -- & -- & -- & -- &  $\sim19$ & $\sim82$ & $6.02^{+2.53}_{-0.97}$\\
           & A3  & $18.12\pm0.01$ &  $19.23\pm0.01$ & $20.28\pm0.01$ &  $21.38\pm0.09$ & 1.93 & $1.05\pm0.08$ & 2.1 & $\sim12$ & $\sim52$ & $0.25^{+0.10}_{-0.04}$ \\
\hline
   RBS-1015 & B1 & -- & $23.79\pm0.04$ & $24.62\pm0.02$ & -- & 0.66 & $0.83\pm0.06$ & 1.9 & $\sim38$ & $\sim145$ & --  \\
\hline
   RBS-1029  & B1 & -- & $22.36\pm0.03$ & $23.12\pm0.03$ & -- & 0.59 & $0.76\pm0.06$ & 1.9 & $\sim37$ & $\sim105$ & -- \\
\hline
   RBS-1460 & B1 & -- & $23.39\pm0.07$ & $24.06\pm0.07$ & -- & 0.65 & $0.67\pm0.15$ & 2.2 &  $\sim42$ &  $\sim148$ & -- \\
\hline
   RBS-1712 & A1 & $22.66\pm0.05$ & -- & $23.29\pm0.02$ & -- & -- & -- & 2.6 & $\sim20$ &$\sim63$ & $0.41^{+0.25}_{-0.04}$  \\
           & A2 & $23.34\pm0.05$ & $24.17\pm0.13$ & $23.70\pm0.02$ & -- & 0.64 & $-0.47\pm0.15$ & 1.9 & $\sim21$ & $\sim66$ & $0.45^{+0.28}_{-0.45}$ \\
\hline
   RBS-1748 & B1$^5$ & -- & $25.09\pm0.05$ & $25.27\pm0.14$ & -- & 0.57 & $0.18\pm0.18$ &  2.2$^5$ & $\sim31$ & $\sim115$ & --\\
   & B2 & -- & $24.78\pm0.04$ & $25.11\pm0.14$ & -- & 0.57 & $0.33\pm0.18$ & 1.9$^5$ & $\sim35$ & $\sim130$ & --\\
\hline
\end{tabular}
\end{center}
}
\end{table}
\end{center}
\end{landscape}

%------------------------------------------------------------------------------------------------------------
\begin{acknowledgements}
This work is supported by the Austrian Science Foundation (FWF)
project number 15868.
\end{acknowledgements}
%\addcontentsline{toc}{chapter}{Bibliography}
\bibliographystyle{aa}
\bibliography{ARCRAIDER_II_arxiv}\label{bibl}
%------------------------------------------------------------------------------------------------------------
\appendix{}\label{app:images}
\section{Images}
%------------------------------------------------------------------------------------------------------------
please download the images from\\
\tt http://astro-staff.uibk.ac.at/$\sim$w.kausch/ARCRAIDER\_II\_images.tar.gz\rm\\

\end{document}